\documentclass[aip,amsmath,amssymb,reprint]{revtex4-1}

\usepackage{graphicx}
\usepackage{dcolumn}
\usepackage{bm}

\usepackage[utf8]{inputenc}
\usepackage[T1]{fontenc}
\usepackage[english]{babel}
\usepackage{mathptmx}
\usepackage{tikz}
\usepackage{amsmath}
\usepackage{etoolbox}
\usepackage{hyperref}
\usepackage{verbatim}
\usepackage{float}
\hypersetup{
    colorlinks=true,
    linkcolor=cyan,
    filecolor=cyan, 
    citecolor=cyan, 
    urlcolor=cyan,
    pdftitle={Overleaf Example},
    pdfpagemode=FullScreen,
    }
\usepackage{caption}
\usepackage{subcaption}
\usepackage[export]{adjustbox}
\makeatletter
\def\@email#1#2{%
 \endgroup
 \patchcmd{\titleblock@produce}
  {\frontmatter@RRAPformat}
  {\frontmatter@RRAPformat{\produce@RRAP{*#1\href{mailto:#2}{#2}}}\frontmatter@RRAPformat}
  {}{}
}
\makeatother
\begin{document}
\title[The Soft SFA]{The Soft-Membrane Surface Forces Apparatus}
\date{\today}
\author{Ilyes Jalisse}
\affiliation{Univ. Bordeaux, CNRS, CRPP, UMR 5031, Pessac F-33600, France}  
\author{Aditya Jha}
\affiliation{Univ. Bordeaux, CNRS, LOMA, UMR 5798, Talence F-33400, France}
\author{Lionel Buisson}
\affiliation{Univ. Bordeaux, CNRS, CRPP, UMR 5031, Pessac F-33600, France} 
\author{Fr\'ed\'eric Nallet}
\affiliation{Univ. Bordeaux, CNRS, CRPP, UMR 5031, Pessac F-33600, France} 
\author{Yacine Amarouchene}
\affiliation{Univ. Bordeaux, CNRS, LOMA, UMR 5798, Talence F-33400, France}
\email{yacine.amarouchene@u-bordeaux.fr}

\author{Thomas Salez}
\affiliation{Univ. Bordeaux, CNRS, LOMA, UMR 5798, Talence F-33400, France}
\email{thomas.salez@cnrs.fr}

\author{Carlos Drummond}
\affiliation{Univ. Bordeaux, CNRS, CRPP, UMR 5031, Pessac F-33600, France} 
\email{carlos.drummond@crpp.cnrs.fr}

\begin{abstract}
Compliant walls are widespread in biological and engineering systems. Because of their singular nature, adapted tools are required to accurately study their rheological properties as well as the consequences of the latter within a given mechanical setting. Because of their slender nature, membranes can be considered as prototypical examples of highly compliant boundaries. In this study, we describe a modified Surface Forces Apparatus (SFA) developed to measure the forces acting on a compliant membrane by measuring its deformation field. We discuss how such a device can be used to characterize the rheology of suspended membranes and accurately measure the electrostatic interactions between a polarized membrane and a spherical electrode, without the need of an external measurement spring. 
\end{abstract}
\maketitle 

\section{\label{1} Introduction}
Surface forces determine the properties of numerous physical systems. They play an essential role in colloidal stability, adhesion, wettability, friction, and many biological processes. Surface-force studies date back to 1928 when Tomlinson investigated the interactions between crossed filaments of different metals.\cite{Tomlinson1928} Later, research groups in the Netherlands and Russia, led by Overbeek\cite{Overbeek1954} and Derjaguin,\cite{Rabinovich1982} developed techniques for measuring the forces between two surfaces of quartz or glass for distance separations above ca. 10~nm. These studies preceded the work of Tabor, Winterton, and Israelachvili, which led to the development of the Surface Forces Apparatus (SFA).\cite{TABORD1969,ISRAELACHVILI1972} In the interferometric version of this technique, the separation between back-silvered molecularly smooth mica surfaces is measured (at nanometric distances, down to the contact between the surfaces), and the interaction force  is determined from the deflection of a double-cantilever spring attached to one of them. 

A more recent and now widespread related technique is the Atomic Force Microscopy (AFM), which measures the interaction between a tip of sub-micrometric size attached to a compliant cantilever and a substrate. In a typical AFM force-distance measurement, the tip-substrate separation is changed at a constant speed while monitoring the cantilever deflection using a light-lever method. The interaction force can be accurately measured as a function of the tip-substrate distance from the spring deflection if the elastic constant of the cantilever is known beforehand. However, the actual separation between the interacting AFM tip and substrate can only be determined in an absolute fashion  if and after tip-substrate contact is reached.\cite{Butt2005}

The practical use of the results from the experimental studies regarding the interaction forces (e.g., to describe the behavior of colloids or to develop a given theoretical description) requires considering the geometry of the interacting surfaces and their deformation under stress in a self-consistent approach. However, such information is not always available. In AFM force measurements, the tip geometry can only be determined ex-situ, and its deformation can only be inferred from contact mechanical models.\cite{Butt2005} A more direct determination of the geometry of the interacting surfaces can be obtained in SFA, where the actual separation between the surfaces and their shapes are measured using white-light interferometry. However, such pieces of information are typically obtained using silver mirrors coated on the back sides of the interacting surfaces. As a consequence, changes occurring on the upper sides of the surfaces (e.g. the deformation of an adsorbed layer) are difficult to measure. A strategy to overcome this limitation, based on the fine analysis of the interference patterns and the contributions of the different layers was proposed by Heuberger and coworkers,\cite{Heuberger1997} but has not been widely implemented.

The need to monitor the interfacial geometry is even more important for systems involving compliant interfaces. For relatively rigid bodies, only deformations close to the interaction zone must be considered, and accurate contact mechanical models have been developed.\cite{Johnson1985} On the contrary, the shape of more compliant solids under stress can change over much larger distances; this fact must be taken into account to properly disentangle the long-range aspect of surface forces.\cite{Attard1992,Barthel1998,Parker1992} In dynamic settings, the interfacial deformation due to hydrodynamic forces can have significant consequences.\cite{leroy2012hydrodynamic,Villey2013,wang2015out} For instance, the emergence of a lift force at zero Reynolds numbers resulting from the combination of confined viscous flows and compliant boundaries\cite{Sekimoto1993,Beaucourt2004,Skotheim2005,Bureau2023,rallabandi2024fluid} has been recently demonstrated experimentally by several groups.\cite{Saintyves2016,Davies2018,Rallabandi2018,Vialar2019,Zhang2020,Fares2024} Such a motion-induced normal force pushes the surfaces apart, hence leading to a significant reduction of effective friction forces, with key implications for industry and physiology for instance.

In a series of studies on interactions involving fluid interfaces, Horn, Chan, and coworkers studied the coupling between surface forces, hydrodynamics, and surface deformation.\cite{Chan2008,Manica2008} In particular, they demonstrated that by monitoring the interfacial deformation, the description of the system dynamics could be envisioned. This description, which combines surface and hydrodynamic forces, takes advantage of the connection between forces and interfacial deformation. Thus, forces can be self-consistently derived from deformation if an appropriate constitutive equation is used. It thus appears that an equivalent approach may be developed for a compliant solid interface, provided that the geometry can be mapped out. Further improvements in this direction would be of clear use for the non-invasive, contactless mechanical characterization of soft, fragile and potentially alive solids (e.g., biological cells).

In the present article, we report on the design and principles of a new generation of SFA: \textit{the Soft SFA}. The latter 
allows to measure static and dynamic forces involving a rigid sphere and a slender, compliant membrane, whose shape can be accurately measured using Multiple Beam Interferometry. Using this device, the membrane rheological properties can be characterized from the response to a well-defined external perturbation. Conversely, we describe how the membrane and its electrostatic and dynamic responses provide fine measurements of the forces at play. 

\section{\label{II} Experimental setup}
\subsection{\label{IIA} Operation principles }

We have developed a modified Surface Forces Apparatus, integrating a compliant membrane as one of the interacting surfaces. The interest of this configuration is twofold. First, as the membrane acts as an accurate force sensor itself, the double-cantilever spring commonly used to measure forces in SFA can be omitted, hence leading to a simpler design. Second, the coupling between elastic deformations and hydrodynamic interactions -- central to many soft-matter settings -- can be studied with the new configuration we propose, which is not possible with a classical rigid SFA.

As commonly implemented in SFA,\cite{Israelachvili1973} we use white-light Multiple Beam Interferometry (MBI) to accurately detect changes in the geometry of a pair formed by a rigid hemisphere and a compliant circular membrane. The schematic of the instrument is shown in Fig.~1. A thin, circular, free-standing, silver-coated PDMS membrane of radius (of the free-standing part only) $a=5.5$~mm,  and thickness $t_{\textrm{m}}$, is mounted on a rigid holder, which is attached to a piezoelectric nanopositioner (P-753 Linear Actuator, Physik Instrumente) for a precise control of the membrane vertical positioning. A silver-coated hemispherical lens (borosilicate Half-Ball Lens, Edmund Optics) of radius $R_0=5$~mm is fixed in front of the silver-coated membrane, separated by a distance $D$. White light passing through the membrane-sphere cavity follows multiple reflections between the two silver mirrors. The wavelengths that undergo constructive interference are transmitted through the top surface, providing information about the shape and optical thickness of the region in between the mirrors. Then, spectral decomposition of the emerging beam in an imaging spectrometer (Jobin Yvon Triax 550) allows for an accurate determination of the sphere-membrane distance $D$ (measured with an accuracy ca. 0.1 nm) and the shape of the membrane. We use the multilayer matrix method for spectral analysis and determination of the system geometry, as has been widely discussed before.\cite{Heuberger2001}

\begin{figure}[ht!]
    \includegraphics[scale = 0.2]{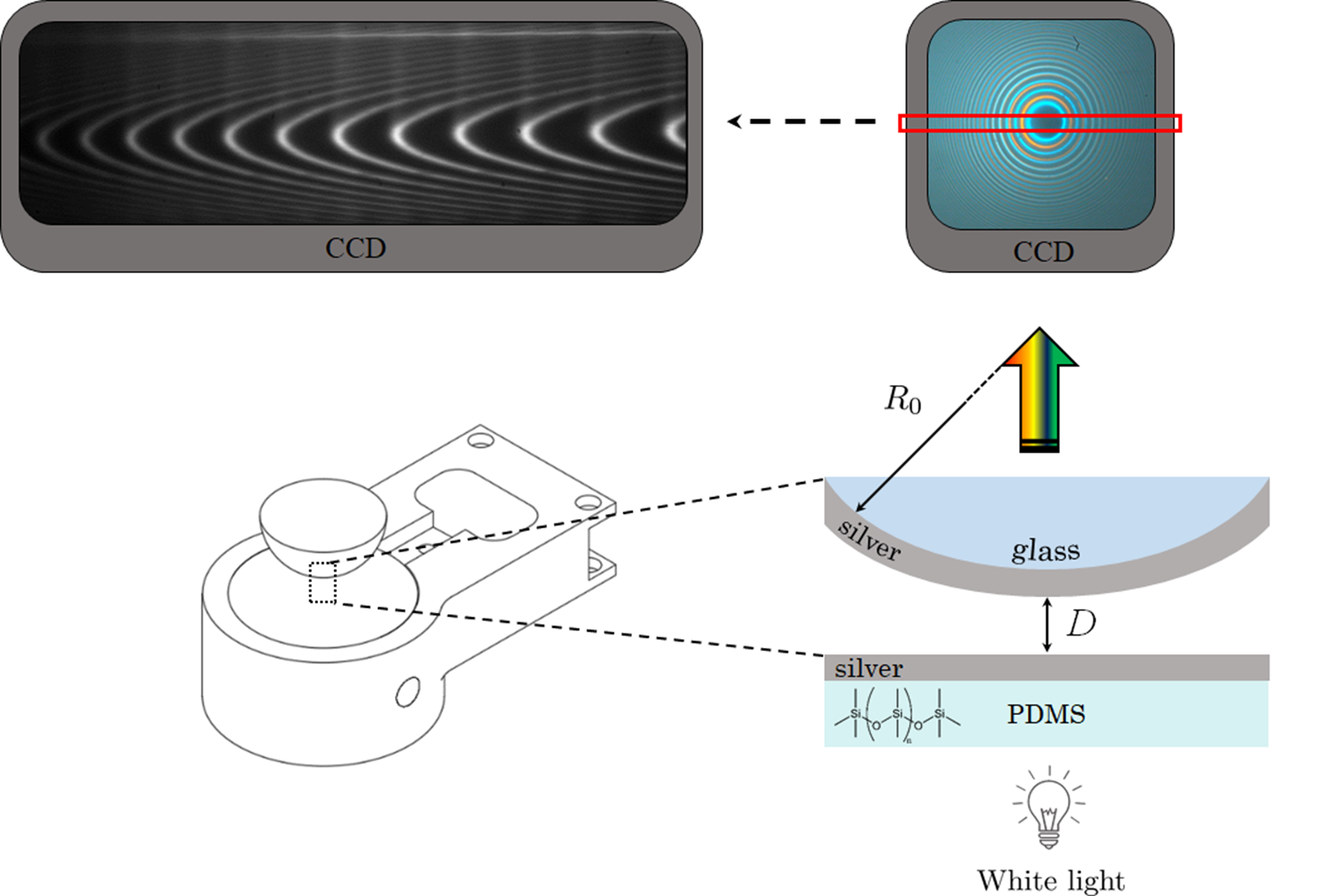}
    \caption{\label{SFA} Schematic of the Soft-Membrane Surface Forces Apparatus (Soft SFA). A silver-coated hemispherical rigid lens of radius $R_0$ is placed in front of a silver-coated elastomeric PDMS membrane, both being separated by a distance $D$. A voltage is applied between the two parts, and white-light interferometry allows for the reconstruction of the gap profile.}
\end{figure}

\subsection{\label{IIB} Thin elastic membrane}
\subsubsection{Membrane preparation}
We prepare free-standing compliant membranes, by using Sylgard 184 (Dow Corning). Sylgard is a poly(dimethylsiloxane) (PDMS)-based  kit that is commonly used in microfluidics for the preparation of elastomers. It is relatively easy to use, transparent, and exhibits good chemical and thermal stabilities. The precise composition of SYLGARD 184 is proprietary. It is often described as a two-component material, involving the base resin, dimethylvinyl-terminated telechelic dimethyl siloxane chains (component A), and a crosslinker (tetramethyl tetravinyl cyclotetrasiloxane) (component B). However, other components are incorporated by the manufacturer into the formulation. In particular, a large concentration of silica-based nanoparticles (dimethylvinylated and trimethylated silica, up to 60$\%$ w/w) is included to improve the elastic properties and resistance to rupture of the resulting material.\cite{Mazurek2019} The PDMS chains are crosslinked after parts A and B are mixed together. The final properties of the elastomer depend on the A:B mixing ratio and the curing thermal history.

\begin{figure}[!h]
     \centering
    \includegraphics[scale = 0.15]{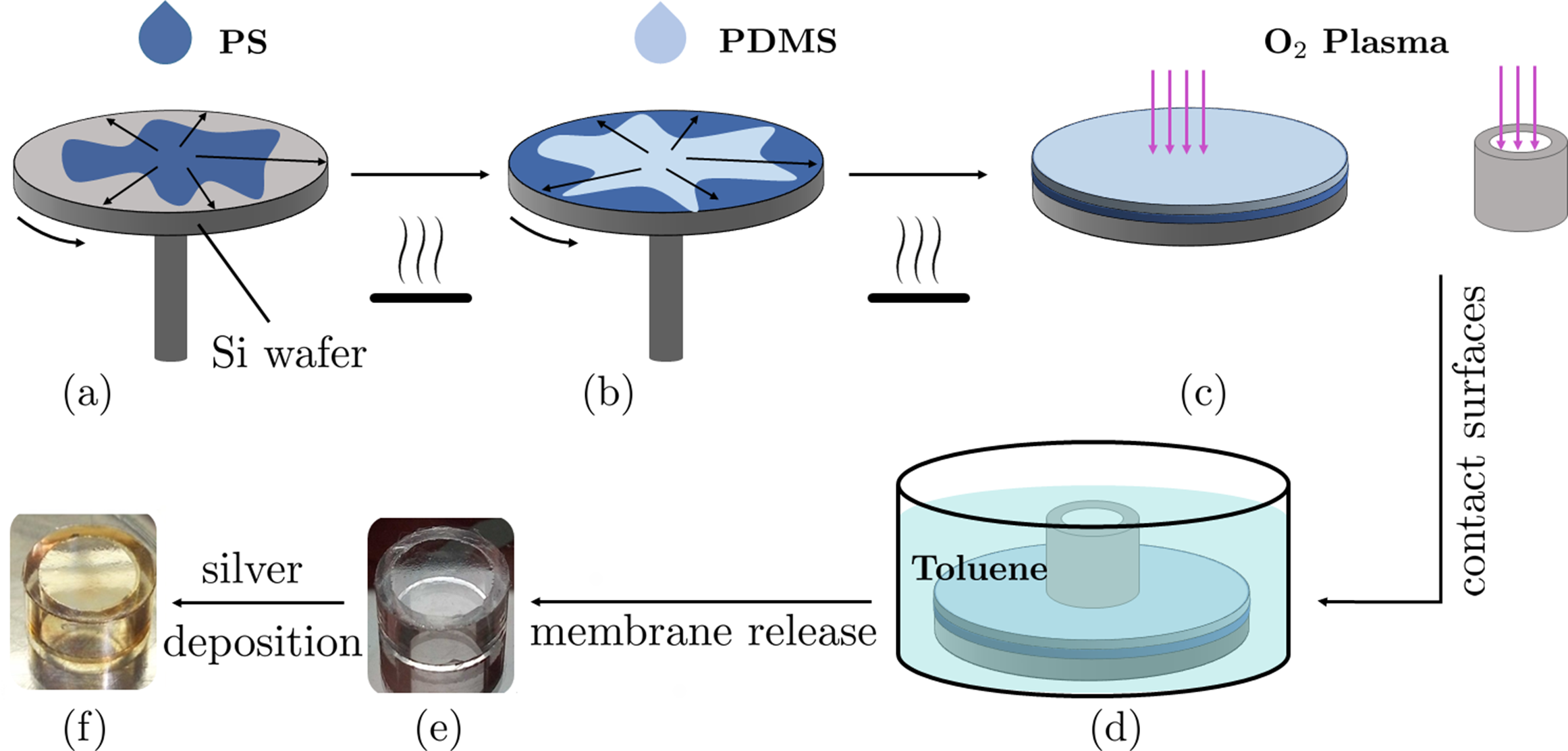}
    \caption{Preparation protocol of the PDMS membranes. (a) Spin coating of a PS solution. (b) Spin coating of PDMS before curing in an oven. (c) $O_2$-plasma  surface activation of both the glass cylinder corona and the PDMS surface. (d) Membrane immersion in toluene to dissolve the PS film. (e) Membrane release. (f) Silver coating of the membrane by evaporation. }
    \label{film preparation}
\end{figure}

Spin coating is used to manufacture thin PDMS films (Fig.~\ref{film preparation}). It allows in particular to achieve an accurate control of their thickness. Briefly, we proceed as follows. A silicon wafer is cleaned for 15 min in a UV-ozone cleaner (UVOCS), rinsed with ethanol to remove impurities, and dried with nitrogen gas. A 50-nm-thick sacrificial layer of polystyrene (PS) is then spin-coated onto the wafer (Laurell WS-650Mz spin coater), using a 2$\%$ PS-toluene solution (Fig.~\ref{film preparation}a). This layer facilitates the detachment of the PDMS film at the end of the process. The wafer-PS sample is heated at 70~°C during 15 min to ensure complete toluene evaporation. Then, we prepare a Sylgard-184 mixture with a precise mass ratio of cross-linking agent to elastomer base, $X$. The mixture is dissolved in hexane (1:1 w/w) to reduce its viscosity and facilitate the spin coating process. The preparation is then mixed during a few minutes for homogenization. After mixing, a film of the mixture is deposited on top of the thin PS film (Fig.~\ref{film preparation}b) and the wafer-PS-PDMS sample is cured at a particular temperature during 2 h. Both the curing temperature and the curing time play key roles in determining the mechanical properties of the final membrane. Once the curing is complete, we use an 8-mm-tall glass cylinder as a support for the free-standing PDMS membrane. This holder is cut from a borosilicate glass tube (11 mm inner diameter and 15 mm outer diameter), and its annular face is sanded to improve the contact with the membrane by decreasing the roughness. The membrane is then irreversibly bonded to the glass cylinder after oxygen plasma surface activation (Quorum, ref. power 50 W for 1 min on glass and PDMS surfaces, Fig.~\ref{film preparation}c). Immediately after plasma treatment, the treated surfaces are brought into contact and cured in an oven at 70$^\circ$C for 1 min to accelerate bonding. The next step is the membrane release. Toluene exposure (Fig.~\ref{film preparation}d) during a few minutes allows to completely dissolving the PS sacrificial film, while the PDMS film remains glued to the glass cylinder. Finally, the top PDMS surface is coated with a 41-nm-thick silver film by thermal evaporation. 

As detailed hereafter, several tests were performed to evaluate the properties of the produced membranes and validate the reliability of the preparation method. In particular, the homogeneity of the membranes can be assessed, and the typical pre-stress related to the preparation procedure can be quantified.

\subsubsection{Static response}
 The static response of a given membrane can be explored by performing a bulge test. This is a widely recognized technique for determining the Young's modulus and the pre-stress of thin films.\cite{Small1992} When subjected to an external pressure $\Delta P$, the thin film exhibits an out-of-plane deflection, which is determined from the intrinsic material properties (Young's modulus E and Poisson's ratio $\nu$), geometrical parameters (film thickness $t_\textrm{m}$ and radius $a$), residual stress $\sigma_0$ (or the associated in-plane tension $N_0$, with $\sigma_0=N_0/t_\textrm{m}$), boundary conditions (e.g. supported vs clamped films), and the applied external load. Typical examples of external loads include uniform hydrostatic pressure or a localized load applied by an indenter. Constitutive equations are required to relate the mechanical response of a system to externally applied forces. Then, converting the force-displacement data into stress-strain data allows for the mechanical characterization of the film, with the extraction of the elastic parameters and the residual stress. Conversely, if the properties of the membrane are known, the film can be used as a sensor and the external force field responsible for the measured membrane deformation can be determined.
 
 The mechanical response of loaded membranes has been widely investigated. Complete descriptions of the behavior of films loaded under different conditions have been reported by Komaragiri et al. \cite{Komaragiri2005} and Wan and Dillard,\cite{Wan2003} although exact analytical expressions that allow the calculation of the applied load from the measured deflection are not available. Nevertheless, accurate  expressions relating load and membrane deformation, based on finite-element calculations, have been reported by several groups. For the case of a membrane under uniform hydrostatic pressure as in a bulge test, Small and Nix\cite{Small1992} showed that the membrane deflection $w_0$ at the center can be well described by:

\begin{equation}
    \Delta P = \dfrac{8Et_\textrm{m}}{3(1-\nu)a^4} (1-0.241\nu)w_0^3+\dfrac{4\sigma_0t_\textrm{m}}{a^2}w_0\ .
 \label{bulge}
 \end{equation}   
   
Similar expressions have also been proposed by other groups.\cite{Pan1990} Wan and Dillard derived equivalent parametric equations for the applied pressure and the membrane deformation in terms of the total membrane stress, which combines residual and stretching-induced stresses.\cite{Wan2003}  Analogous expressions have been proposed for different varieties of the applied load (e.g. localized load).\cite{Wan2003,Vinci1996} Equation (\ref{bulge}) involves the balance between the applied pressure and the total elastic stress. As such, a simple fit of the measured relation between $\Delta P/w_0$ and $w_0$ allows for the determination of both the pre-stress and the elastic  modulus, provided that the other relevant parameters are known. At low pressures (i.e. small deformations), the membrane response is dominated by the pre-stress. On the contrary, at large deformations, the effect of the pre-stress is negligible compared with the externally applied stress. 

The bulge-test setup is illustrated in Fig.~3a. The glass cylinder supporting the membrane is placed in a sealed aluminum holder. The setup consists of an inlet, connected to a water-filled syringe, which allows for water to fill the chamber underneath the membrane. For every experiment, we ensure that air is not present in the reservoir. The hydrostatic pressure acting on the membrane is carefully controlled by varying the height of the water column inside the syringe. The membrane deflection is simultaneously monitored using a laser-displacement sensor (Keyence LK-H008) mounted on a fixed frame above the freestanding membrane. The device can measure vertical displacements of up to 500 µm with a precision of 1 µm. 

\begin{figure}[ht!]
    \begin{subfigure}[c]{.2\textwidth}
      \includegraphics[width=0.98\columnwidth,left]{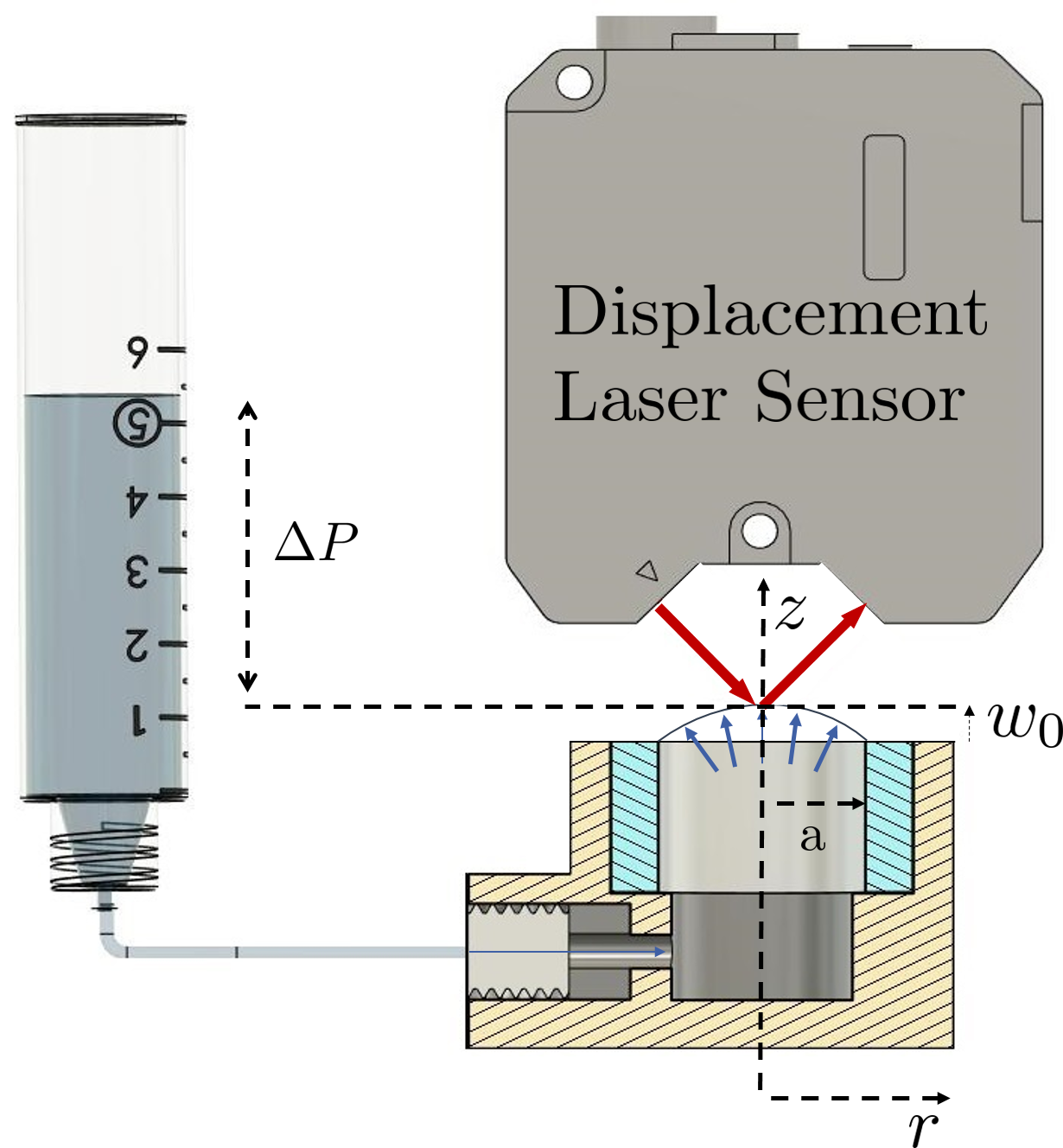}
      \caption{}
      \label{bulge_test_experiment}
    \end{subfigure}
        \hspace{1cm}
    \begin{subfigure}[c]{.2\textwidth}                     \includegraphics[width=1.28\columnwidth,right]{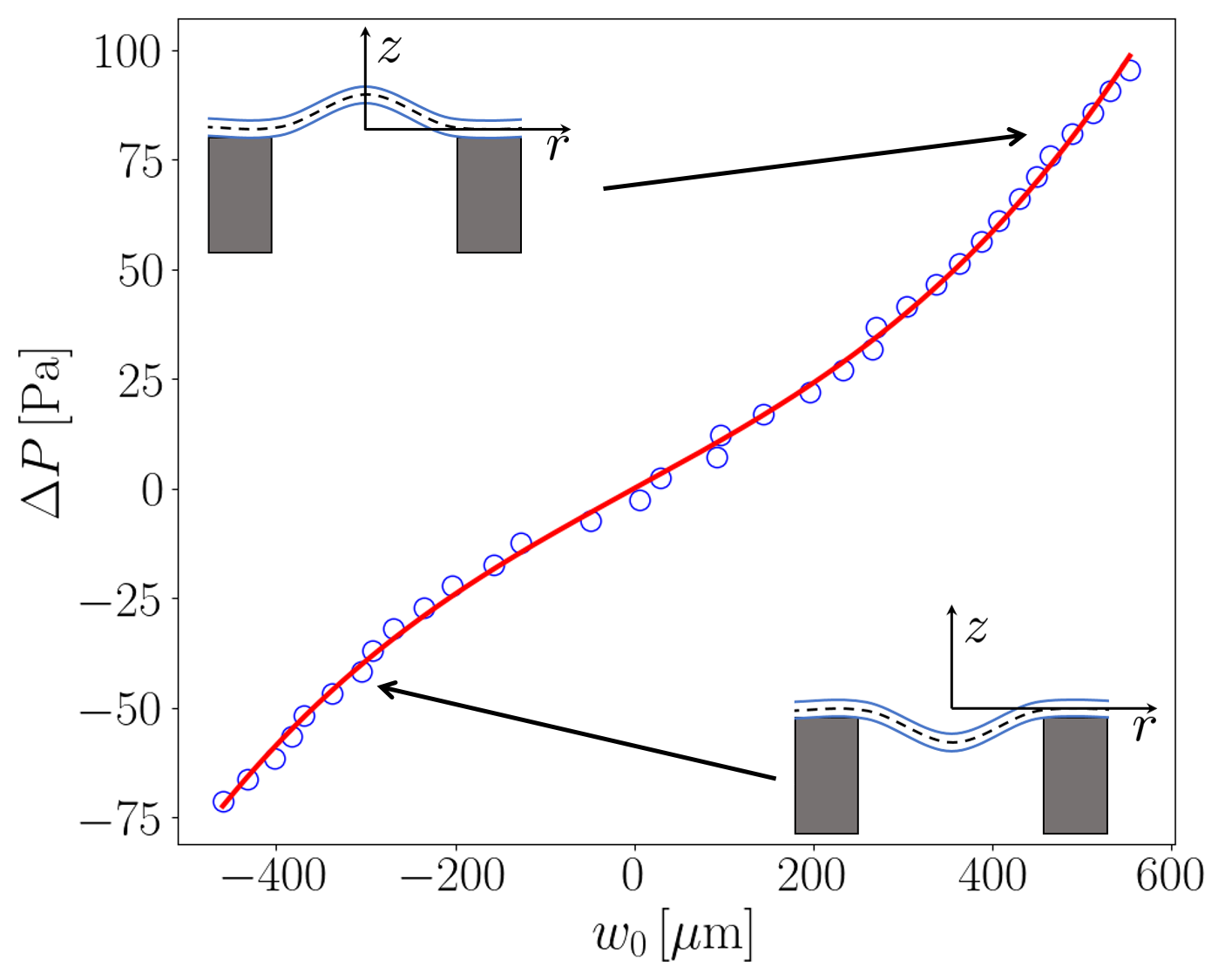}
      \caption{}
      \label{theory_fit} 
    \end{subfigure}
    \caption{Hydraulic bulge test. (a) Schematic of the setup. A water-filled syringe placed at a controlled height is connected through a silicone tube to a sealed chamber containing the free standing membrane. The membrane deflection is monitored  using  a  laser-displacement sensor. (b) Typical pressure-deflection data, fitted to Eq.~(\ref{bulge}) by fixing the membrane thickness  $t_\textrm{m} = 15.4$~µm, membrane radius $a = 5.5$~mm, and membrane Poisson's ratio $\nu=0.5$. The two fitting parameters are the pre-stress $\sigma_0 = 54.8$~kPa and the membrane Young's modulus $E = 2.8$~MPa. Curing temperature 85$^\circ$C. $X=1{:}5$.}
\end{figure}

Typical bulge-test results are shown in Fig.~\ref{theory_fit}. The thickness $t_\textrm{m}$ of the PDMS membrane is independently measured using MBI.  Fitting the experimental bulge-test data to Eq.~(\ref{bulge}), we find that $\sigma_0= 54.8$~kPa and $E=2.8$~MPa. These values strongly depend on the fabrication procedure of the membrane, particularly with respect to curing time and temperature.

\subsubsection{Dynamic response}
The vibrational modes of circular elastic membranes are well known.~\cite{Wah1962,wah1963vibration,amabili2008nonlinear,PeterHowellJohnOckendon2009} In particular, when the bending contribution can be ignored, the modal frequencies of a stretched ideal membrane are given by\cite{Wah1962}:
\begin{equation}
    \Delta f_{ij} = \dfrac{\mu_{ij}}{2\pi a} \sqrt{\dfrac{N_0}{\rho t_{\textrm{m}}}} \ ,
 \label{resonances}
 \end{equation}   
where $\rho$ is the mass density of the membrane, $\mu_{ij}$ is the $j$-th root of the order-$i$ Bessel function $J_i$, and $N_0$ is the radial stress resultant (with unit of a force per unit length for a membrane) at the clamping boundary ($r=a$), i.e. the in-plane tension in the membrane. By using Eq.~(\ref{resonances}), the in-plane pre-stress $\sigma_0=N_0/t_{\textrm{m}}$ can be evaluated from the measured resonance frequencies, provided that the orders $i$ and $j$ of the vibrational modes can be identified. Different techniques have been proposed to characterize the resonance frequencies of a clamped membrane, with variations in the means of excitation and methods of measurement of the deformation. In this work, we have combined: i) Electronic Speckle Pattern Interferometry (ESPI) of the membranes, excited with a sinusoidal pressure signal; ii) spectral analysis of the thermal motion of the same membranes. We present both methods hereafter.

ESPI is a non-destructive whole-field technique that allows the visualization of the out-of-plane displacement of a vibrating object with high resolution.\cite{Maden1994,Moore2004} This is achieved by determining the difference in optical paths between two speckle patterns resulting from the interference between a reference laser beam and a reflected laser beam, at different excitation levels.\cite{Moore2004,yang_review_2014} The experimental setup for the ESPI is depicted in Fig.~\ref{manip_holo}. The light source is a 200 mW and 532 nm NdYAG green laser (Spectra-Physics Excelsior). The power range is controlled by passing it through a half-wave plate and a polarizer. The laser is injected into a single-mode fiber at 532 nm Y 50/50 (OZ Optics) using a fiber coupler (KT120, Thorlabs). At the fiber exit of each arm, the beam is collimated using a coupler and then expanded (GEB05, Thorlabs). The size of the beam is adjusted to match the size of the membrane. Then the two beams are recombined by a semi-reflecting mirror (CM1-BS1, Thorlabs) towards the camera  (Lumenera, Lt425C-SCI, $2048$x$2048$ px, 12 bits) fitted with a zoom lens (Edmund  Optics, 0.36X Telecentric Lens).
Image processing is carried out using a custom-made LabVIEW program. We use  loudspeakers fed with a purely sinusoidal input to excite the motion of the membranes, sweeping the excitation frequency  while recording the speckle pattern for both the excited (phase-shifted) and unperturbed (reference) conditions. The subtraction $I(\phi) - I(\phi + \Delta \phi) $ of the light intensity  allows for the visualization of the deformed shape of the membrane (Fig.~\ref{modal shape}) and the identification of the vibrational mode.

\begin{figure}[ht!]
    \include{ESPI_setup}
    \caption{Schematic of the Electronic Speckle Pattern Interferometry (ESPI) setup. A green laser with a 532-nm wavelength is transmitted through a half-wave plate ($\lambda/2$) and a polarizer (P1) before being directed into a single-mode fiber using two mirrors (M1, M2) and a lens (L1). The light power is equally distributed between two other fibers ended by collimated beam expanders (BE). The first one (left) is set to be the reference beam and illuminates directly the CCD camera, while the second one (top) gets reflected onto the metallic surface of the sample membrane before reaching the camera thanks to a beam splitter (BS).}
    \label{manip_holo}
\end{figure}
\begin{figure}[ht!]
    \includegraphics[scale = 0.4]{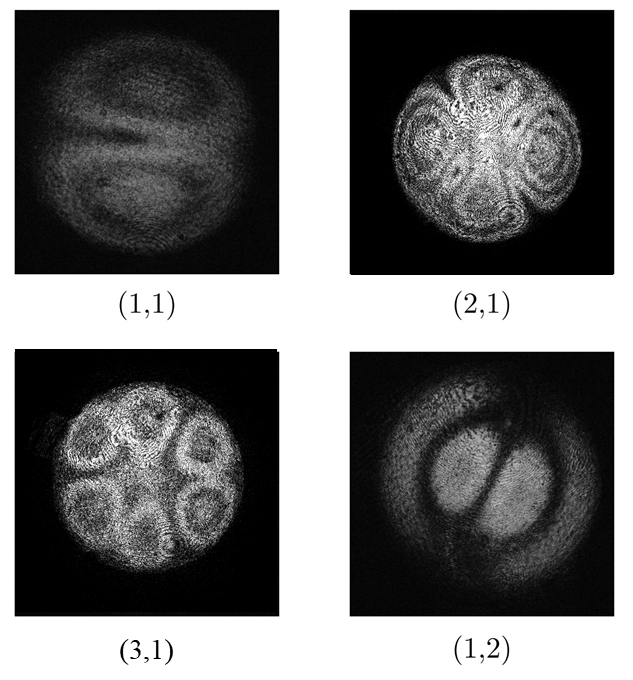}
    \caption{Typical mode patterns of a freestanding PDMS membrane observed using ESPI, along with the $(i,j)$ orders identified. $t_m$ = 14.1 $\mu$m, curing temperature $T = 85^\circ$C. $X=1{:}5$ }
    \label{modal shape}
\end{figure}

\begin{figure}[ht!]
    \includegraphics[scale = 0.3]{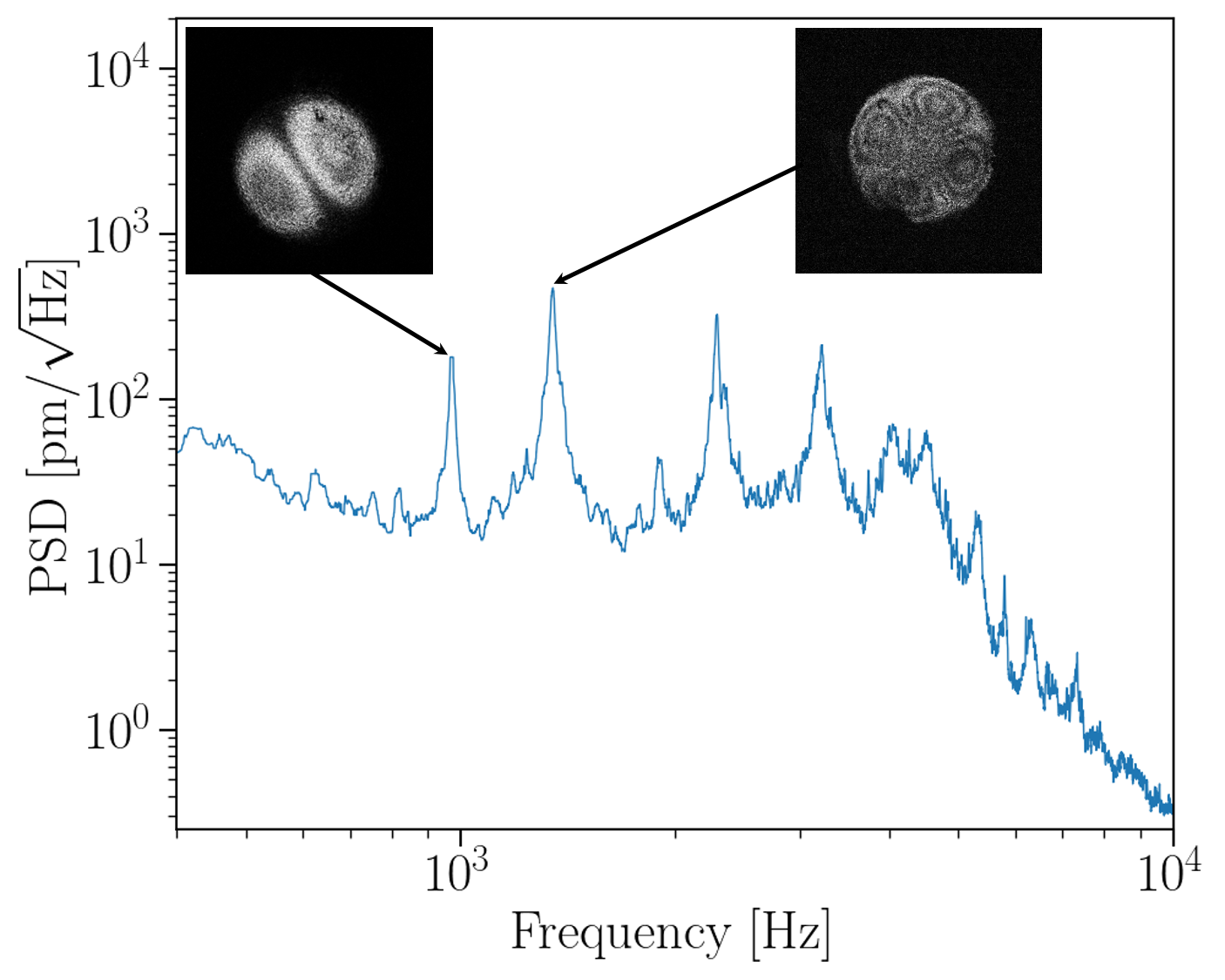}
    \caption{Typical thermal spectrum of a free-standing PDMS membrane with a resolution of 0.6 Hz, measured with a Picoscale interferometer.$t_m$ = 16.9 $\mu$m, curing temperature $T = 100^\circ$C. $X=1{:}5$. Insets: Two ESPI images. Left, mode (1,1). Right, mode (2,1)} 
    \label{vibrationspectrum}
\end{figure}

As shown in Fig.~\ref{modal shape}, the quality, regularity and symmetry of the ESPI images  indicate a good level of homogeneity of the membranes. However, the accurate determination of the resonance frequencies requires a fine monitoring of the system response, which is cumbersome with ESPI. For more detailed spectral characterization, we thus use a Picoscale interferometer. This device is a fiber-optic Michelson interferometer that can be used for contactless displacement measurements with picometer accuracy. Signals are generated in fiber-coupled sensor heads that are connected to a controller via optical fibers. A beam splitter divides the laser beam into a reference beam and a measurement beam, with the reference beam reflected off a mirror (included in the sensor head), and the measurement beam reflected off the target. The interferometer uses a 1550-nm laser (class 1, 400 µW) stabilized by a gas-absorption cell. This setup enables precise quadrature detection through advanced modulation of the native wavelength of the laser. Additionally, the system is integrated with an Environmental Sensor Module that compensates for changes in the air refractive index due to variations in pressure, temperature, and relative humidity. Using this device, we measured the thermal spectra of PDMS membranes in order to accurately determine their resonance frequencies. A typical spectrum is shown in Fig.~\ref{vibrationspectrum}. We can identify all the vibrational modes and relate them to the ESPI data. Therefore, the membrane pre-stress can be computed from the measured resonance frequencies, using Eq.~(\ref{resonances}).

We investigated more than 20 different membranes and consistently observed pre-stress values between 35 and 120 kPa. However, we are not able to correlate such values with the preparation parameters (membrane thickness, curing temperature, or cross-linker-to-base ratio). We can further estimate the strain $\epsilon$ corresponding to the measured pre-stress using the relation $\sigma_0=E\epsilon/(1-\nu)$. Assuming a Poisson's ratio of $\nu=0.5$, values of $\epsilon$ in between 0.03 and 0.04 can be estimated. We believe that the main factor responsible for the observed pre-stress is the difference in the coefficients of thermal expansion between the borosilicate glass (ca. 3 ppm/K), the silicon wafers (2.5 ppm/K) and the PDMS (ca. 250 ppm/K).\cite{Muller2019} The variability in the experimental conditions during curing probably leads to further variation in the pre-stress. The adhesion between the PDMS films and the silicon wafers at early stages of the preparation, and the bonding between the PDMS membranes and the glass cylinder later on severely hinder the relaxation of the membrane after the thermal treatments during cross-linking and bonding.

\section{\label{III} Electrostatic actuation}
\subsection{Experiments}
\begin{figure}[ht!]
    \centering
    \includegraphics[width=0.35\textwidth]{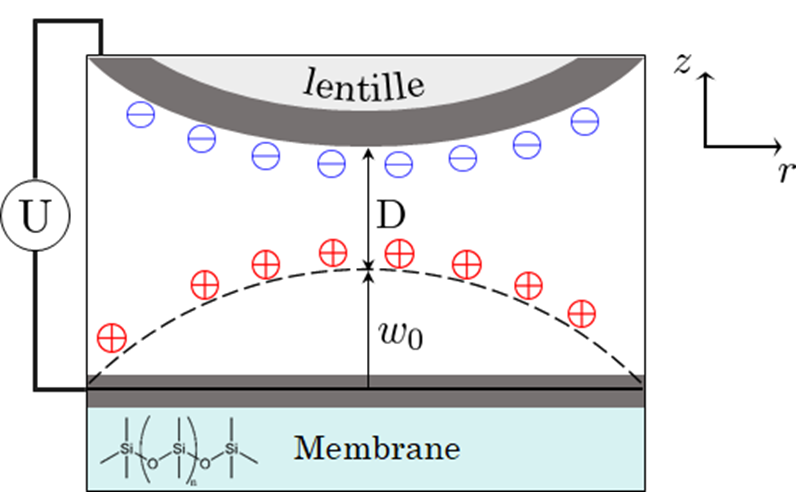}
    \caption{Schematic showing the deflection $w_0$ and the surface charges of the deformed (dashed line) freestanding PDMS membrane, when submitted to a DC voltage $U$ with respect to the rigid hemisphere counterpart of the Soft SFA. }
    \label{deflection_schema}
\end{figure}

We evaluate the performance of the Soft SFA by measuring the response to an externally-applied potential difference between the silver layers on the sphere and the membrane, which can be readily used as electrodes. The  oppositely charged surfaces then attract each other because of Coulombic interactions. Specifically, we apply a Direct Current (DC) voltage $U$ to bias the electrodes (power supply ELEKTRO-AUTOMATIK EA-PS8160-04-T) and measure the deformation $w_0$ at the center of the compliant membrane, as illustrated in Fig.~\ref{deflection_schema}.

\begin{figure}[!h]
    \centering
    \begin{subfigure}{0.47\linewidth} 
        \centering
        \includegraphics[width=\linewidth]{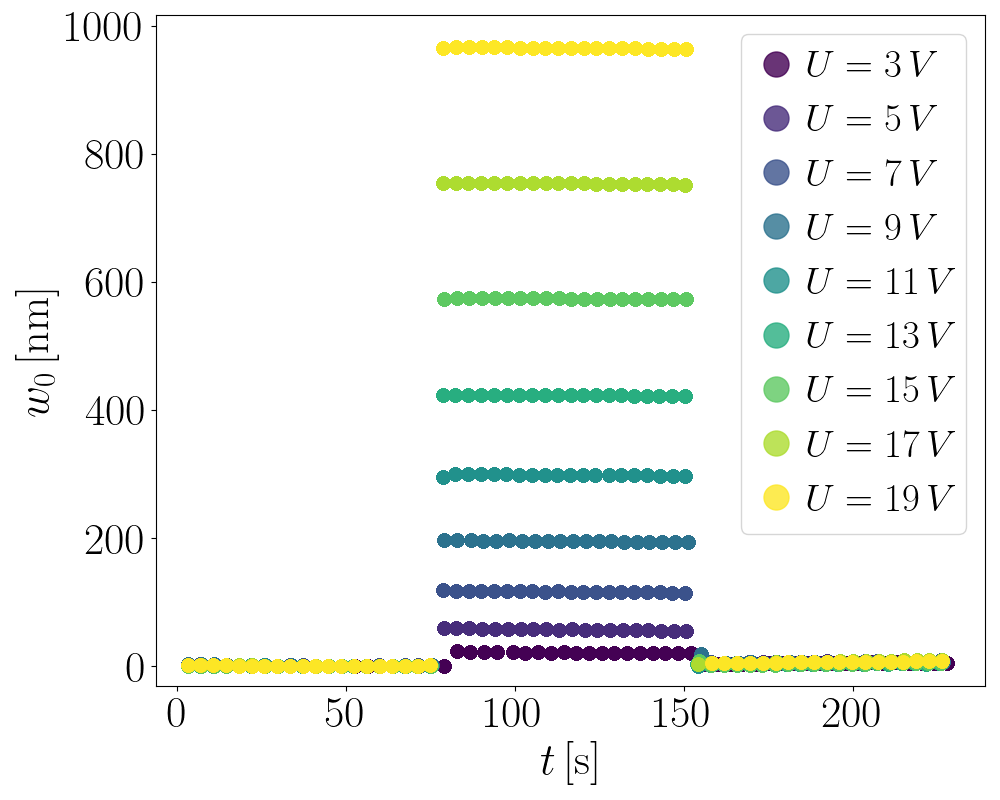}
        \caption{}
        \label{electrostaticresponse_élastique}
    \end{subfigure}
    \hspace{0.02\linewidth} 
    \begin{subfigure}{0.47\linewidth} 
        \centering
        \includegraphics[width=\linewidth]{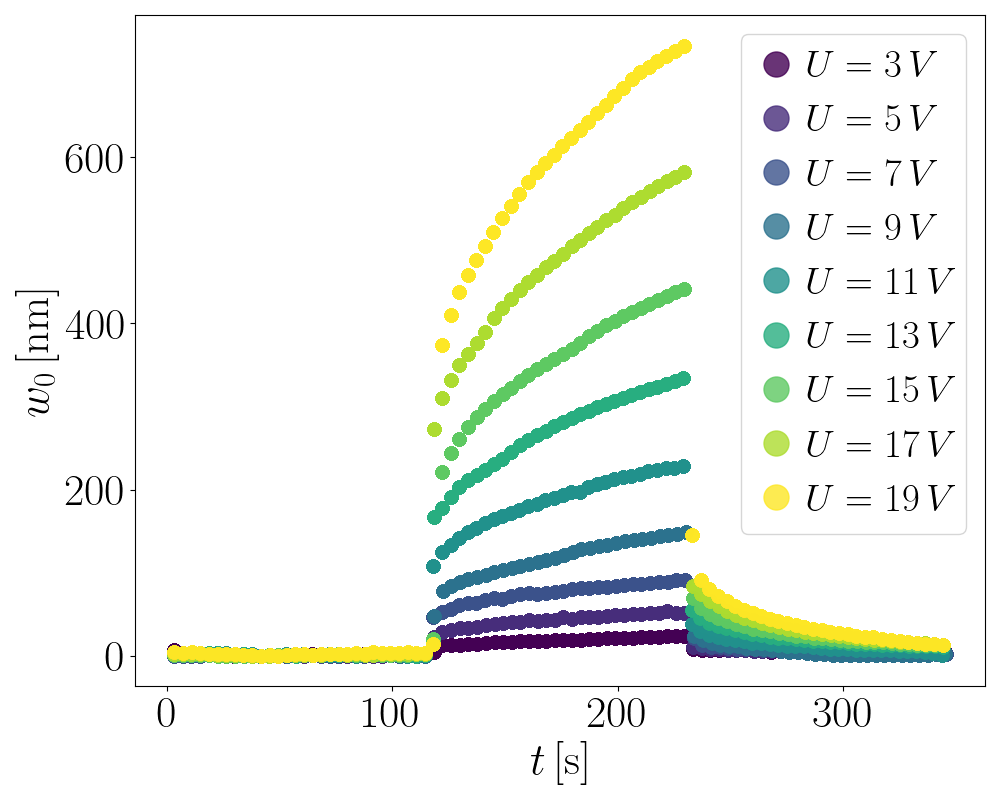}
        \caption{}
        \label{electrostaticresponse_visqueux}
    \end{subfigure}
    \caption{Central deflection $w_0$ of the membrane as a function of time $t$, for different voltages $U$ as indicated, and for two different cross-linking temperatures during the membrane preparation step: a) 100$^\circ$C; and b) 85$^\circ$C. In all cases, the voltage is first off ($U=0$), then turned on to the chosen $U$ value indicated in the legends, and then turned off again. a) $t_m$ = 20.3 $\mu$m. $X=1{:}5$. Initial gap $D = 9000~nm$. b) $t_m$ = 14.1 $\mu$m. $X=1{:}5$. Initial gap $D = 9050~nm$.}
    \label{electrostaticresponse}
\end{figure}

Typical results are presented in Fig.~\ref{electrostaticresponse}, for various voltages and for two cross-linking temperatures during the membrane preparation protocol. In all cases, a quick response of the membrane is detected upon the application or removal of the DC voltage. For a membrane cross-linked at 100~$^\circ$C (Fig.~\ref{electrostaticresponse_élastique}), a stationary deformation state is achieved shortly after the application or removal of the voltage. 
Moreover, the changes in the membrane deflection are reproducible (not shown) and reversible. In sharp contrast, for a membrane cross-linked at 85~$^\circ$C (Fig.~\ref{electrostaticresponse_visqueux}), a more complex behavior is observed. Mainly, a slow viscoelastic temporal evolution of the membrane deformation is observed after any voltage change, and the stationary state is still not reached after two minutes. This observation indicates that the viscoelastic rheology of the membrane can be simply adjusted by changing the curing temperature during the preparation step. Furthermore, the behavior observed in Fig.~\ref{electrostaticresponse_visqueux} is reversible, as: i) the membrane eventually returns to the initial non-deformed state after sufficient time under zero voltage (not shown); and ii) subsequent on-off cycles produce identical responses (not shown). 

Interestingly, the fast elastic response observed upon turning off the voltage is substantially larger in magnitude than the fast elastic response observed upon turning on the voltage (Fig.~\ref{electrostaticresponse_visqueux}). This feature may suggest a strain-softening behavior, where the membrane would become more compliant under strain, and may be related to the Payne effect.\cite{Payne1962,Mazurek2019} The latter effect designates the reduction of storage modulus of particle-loaded elastomers under strain, which is typically reported for strain values above 0.1~$\%$. Here, we may observe a strain-softening effect for much smaller strains (ca. 10$^{-5}$), a regime that is out of reach with common rheometers or Dynamic Mechanical Analyzers (DMA). 

\begin{figure}[ht!]
    \centering
    \includegraphics[width=0.4\textwidth]{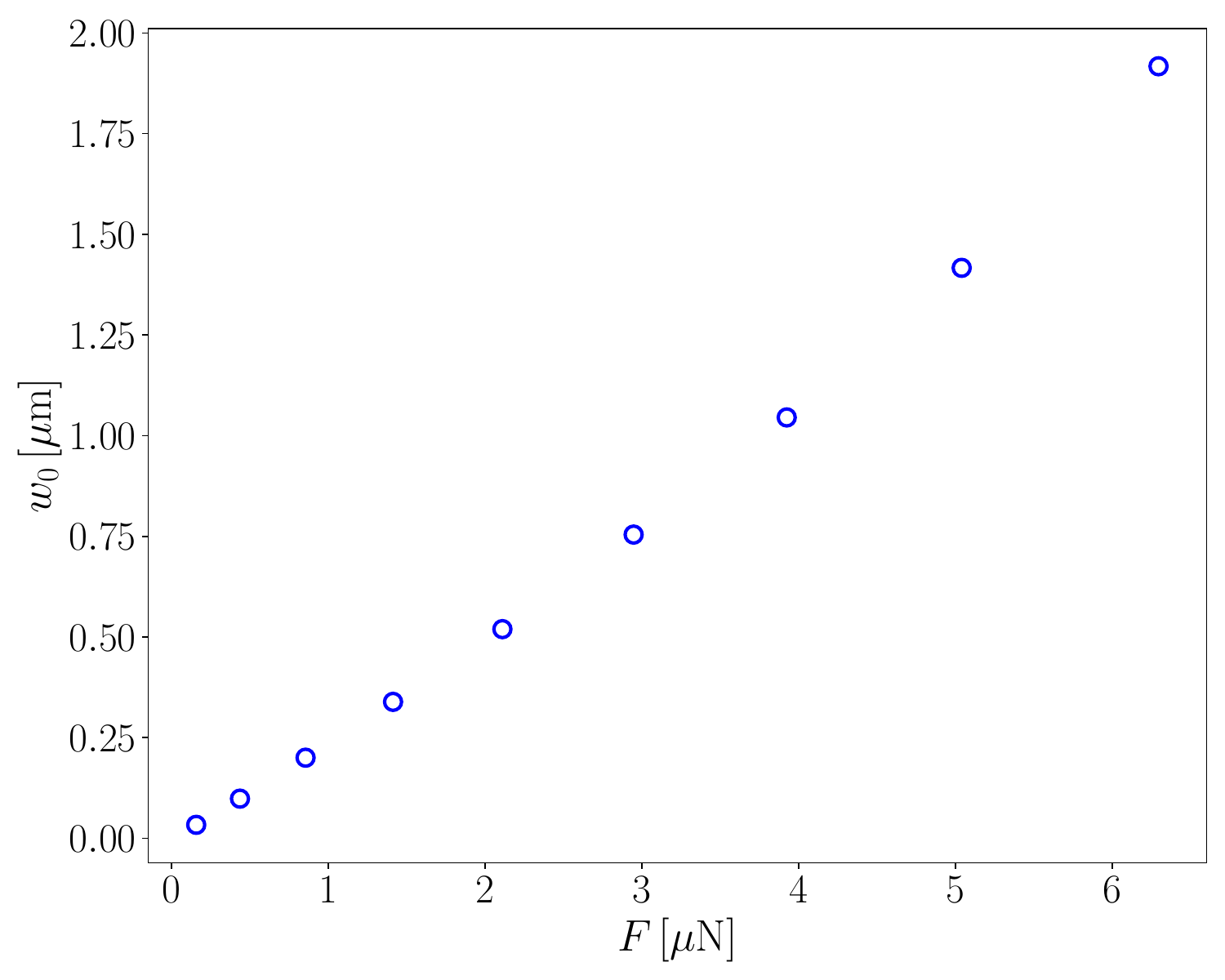}
    \caption{Central deflection $w_0$ of the membrane as a function of the applied electrostatic force $F$, calculated using Eq.~(\ref{eforce}). $t_m$ = 18 $\mu$m, membrane preparation temperature T = 100$^\circ$C.
    $X=1{:}5$.}
    \label{deltavsF}
\end{figure}

We can now assess the detection limit of the force measurement in the Soft SFA. The following approximate expression for the electrostatic force $F$ between a conducting sphere and a conducting plate under a constant potential difference has been  reported\cite{Crowley2008,Lekner2012} 
\begin{equation}
    F = \pi\epsilon_0U^2\dfrac{1}{\xi+\xi^2} 
 \label{eforce}
 \end{equation}  
where $\varepsilon_0$ denotes the permittivity of the gap region, and $\xi=D/R_0$. The deflection at the center of the membrane as a function of the applied electrostatic force $F$ estimated using this expression is presented in Fig.~\ref{deltavsF}. By extrapolation from these data, a membrane deflection of $w_0=1$~nm, which is  above the detection limit of  MBI (0.1 nm), corresponds to a force of $F=4$~nN. Therefore, a resolution $F/R_0$ on the order of 1~µNm$^{-1}$ is readily accessible with the setup described in this work. This is one order of magnitude smaller of what is commonly achieved with conventional SFA setups, highlighting the great potential of the Soft SFA towards force sensing.
\subsection {Theoretical description}
In this section, we provide a theoretical description of the axisymmetric shape of a pre-stressed elastic membrane in an applied electrostatic field.

\subsubsection{Elastic equation}
The out-of-plane deflection field $w(r)$ of the membrane is governed by the F\"oppl-von K\'arm\'an (FvK) equations,\cite{King2012,Smith2022} as follows: 
\begin{align}
	    B\frac{\textrm{d}^{4}w}{\textrm{d}r^4} -\left[(\sigma_0+\sigma_{r})t_\textrm{m}\frac{\textrm{d}^{2}w}{\textrm{d}r^2}+\frac{(\sigma_0+\sigma_{\theta})t_\textrm{m}}{r}\frac{\textrm{d}w}{\textrm{d}r}\right] &= p~, \label{eq:normal_equilibrium} \\
	    \frac{\textrm{d} \sigma_{r}}{\textrm{d}  r}+\frac{1}{r}(\sigma_r-\sigma_{\theta}) &= 0 ~, \label{eq:radial_equilibrium} \
\end{align}
with $p$ being the pressure resulting from the electrostatic interaction between the membrane and the hemispherical probe of radius $R_0$, the in-plane stress fields in the radial ($r$) and azimuthal ($\theta$) directions are respectively denoted by $\sigma_r(r)$ and $\sigma_{\theta}(r)$, and the mechanical properties of the membrane are characterized by its bending stiffness $B=Et_\textrm{m}^3/[12(1-\nu^2$)]. Besides, the central deflection of the membrane is given by $w_0=w(0)$.

We then consider a more specific situation, in which the deformation of the membrane is controlled by the pre-stress $\sigma_0$ only. We can make an order-of-magnitude estimation of the contributions of the different terms in Eq.~(\ref{eq:normal_equilibrium}) to support this approximation. The contribution of the bending term is on the order of $\frac{B w_0}{L^4}$, whereas the contribution of the stretching term is on the order of $\frac{\sigma_0t_\textrm{m}w_0}{L^2}$, with $L$ a typical horizontal length scale in the problem that should be comparable to $a$ in magnitude. As a consequence, and given the membrane parameters and the typical values of the pre-stress measured above, the bending-to-stretching ratio is much smaller than 1, and bending can be neglected. Suppressing further the nonlinear terms within a small-deformation approximation, Eq.~(3) can be rewritten as: 
\begin{align}
	    N_0\frac{\textrm{d}^2 w}{\textrm{d}r^2}+\frac{N_0}{r}\frac{\textrm{d}w}{\textrm{d}r}&= -p~. \label{eq:normal_equilibrium2} 
\end{align}
Finally, since the membrane is clamped on the supporting cylinder at its edges, and due to the axisymmetry of the problem, the boundary conditions are assumed to be: 
\begin{align}
	w(a) = 0~,\\
	\left.\frac{\textrm{d}w}{\textrm{d}r}\right|_{r=0}=0 ~.
\end{align}

\subsubsection{Electrostatic equation}
Invoking the Maxwell-Gauss equation in the absence of bulk charge sources, the electrostatic potential $\psi(r,z)$ within the membrane-probe gap region satisfies the Laplace equation: 
\begin{align}
\nabla^2\psi = 0 ~,
\end{align}
where $\nabla$ is the nabla operator. The hemispherical probe is assumed to be held at a constant electrostatic potential $U$, while the membrane is held at a null potential. We further assume that $U$ is small enough for the membrane to be weakly deformed, i.e. $w_0\ll D$, allowing us to write the profile $h(r)$ of the gap region between the probe and the membrane as the undeformed one: 
\begin{align}
	h(r) = R_0+D-\sqrt{R_0^2-r^2}~. 
	\label{eq:geometry} 
\end{align}
Considering further the small-gap limit, where $\varepsilon = D/R_0 \ll 1$, the latter equation can be approximated by its parabolic expansion: 
\begin{align}
	h(r) = D+\frac{r^2}{2R_0}~. 
\end{align}
Interestingly, the small-gap limit allows for a scale separation between: i) an inner region, for $r\ll\sqrt{2DR_0}$, where the electrostatic loading is dominant; and ii) an outer region, for $r\gg\sqrt{2DR_0}$, where the membrane relaxes freely towards its clamping boundary condition.

Let us now focus on the inner region, and non-dimensionalize the spatial coordinates in the problem by the probe radius $R_0$ of the membrane, the electrostatic potential by the voltage $U$, and the pressure by a pressure scale $p^*$. We denote the resulting dimensionless quantities with an overbar. Therefore, the Laplace equation becomes:
\begin{align}
	\frac{1}{\bar{r}}\frac{\partial }{\partial \bar{r}}\left(\bar{r}\frac{\partial \bar{\psi}}{\partial \bar{r}}\right)+\frac{\partial^2 \bar{\psi}}{\partial \bar{z}^2} = 0~,
\end{align}
while the profile of the gap region becomes:
\begin{align}
	\bar{h} = \varepsilon+\frac{\bar{r}^2}{2}~. 
\end{align}

As previously done~\cite{jeffrey1978temperature,jeffrey1980electrostatics}, we now introduce the following stretched coordinates for the inner region: 
\begin{align}
	Z = \frac{\bar{z}}{\varepsilon},\quad R = \frac{\bar{r}}{\sqrt{2\varepsilon}}~. 
\end{align} 
Thus, the height function $H=\bar{h}/\epsilon$ in the stretched-coordinate system reads: 
\begin{align}
	H = 1+R^2~.
\end{align}
Similarly, the potential $\Psi(R,Z)=\psi(\bar{r},\bar{z})$ in the stretched-coordinate system satisfies:
\begin{align}
	\frac{1}{2\varepsilon}\frac{1}{R}\frac{\partial }{\partial R}\left(R\frac{\partial \Psi}{\partial R}\right)+\frac{1}{\varepsilon^2}\frac{\partial^2 \Psi}{\partial Z^2} = 0~,
\end{align}
which reduces, at leading order in $\epsilon$, to: 
\begin{align}
	\frac{\partial^2 \bar{\psi}}{\partial Z^2} = 0~.
\end{align}
The latter equation is subject to the following boundary conditions: i) at $Z = H, \Psi = 1$; and ii) at $Z = 0$, $\Psi = 0$. The solution reads:
\begin{align}
	\Psi = \frac{Z}{H}~. 
\end{align}
The electrostatic pressure exerted on the membrane reads $p=\epsilon_0(\nabla \psi)^2/2$. Hence, in dimensionless variables where $P(R,Z)=\bar{p}(\bar{r},\bar{z})$, one has:
\begin{align}
	P = \left[\frac{\varepsilon}{2}\left(\frac{\partial\Psi}{\partial R}\right)^2+\left(\frac{\partial \Psi}{\partial Z}\right)^2\right] ~,
\end{align}
provided that we set the pressure scale as $p^* = \epsilon_0U^2/(2R_0^2\varepsilon^2)$.
At leading order in $\epsilon$, one thus gets:
\begin{align}
	P = \frac{1}{(1+R^2)^2}~.
\end{align}
This electrostatic pressure field of the inner region vanishes rapidly as $R\rightarrow+\infty$, which is consistent with the fact that we neglect the electrostatic pressure in the outer region.

\subsubsection{Deflection field}
We consider the deflection field $w_{\textrm{in}}(r)$ in the inner region. By introducing the deflection scale $w^*$, so that we can switch to dimensionless variables through $w_{\textrm{in}}(r)=w^*\bar{w}_{\textrm{in}}(\bar{r})$, and the stretched-coordinate notation $W_{\textrm{in}}(R)=\bar{w}_{\textrm{in}}(\bar{r})$, Eq.~(\ref{eq:normal_equilibrium2}) can be non-dimensionalized as: 
\begin{align}
		\frac{\textrm{d}^2W_{\textrm{in}}}{\textrm{d} R^2}+\frac{1}{R}\frac{\textrm{d} W_{\textrm{in}}}{\textrm{d} R} = -P~,
		\label{innerODE}
\end{align}
provided that we set the deflection scale to be $w^*=\frac{\epsilon_0U^2}{\varepsilon N_0}$. 

Besides, the dimensionless electrostatic pressure $\bar{p}$ (or equivalently $P$) typically decays as $\sim\epsilon^2/\bar{r}^{\,4}$ in the far field, as can be estimated from the inner pressure solution found above. Therefore, at leading order in $\epsilon$, the differential equation for the dimensionless deflection field $\bar{w}_{\textrm{out}}(\bar{r})=w_{\textrm{out}}(r)/w^*$ in the outer region reads:
\begin{align}
\frac{\textrm{d}^2\bar{w}_{\textrm{out}}}{\textrm{d} \bar{r}^2}+\frac{1}{\bar{r}}\frac{\textrm{d}\bar{w}_{\textrm{out}}}{\textrm{d} \bar{r}}= 0~. 
\end{align}
The general solution of the latter equation is $\bar{w}_{\textrm{out}} = c_1\ln(\bar{r})+c_2$, where $c_1$ and $c_2$ are two unknown constants. Since $\bar{w}_{\textrm{out}}(\bar{r}=a/R_0)=0$ from the clamping boundary condition, $c_2$ can be found, and one gets $\bar{w}_{\textrm{out}} = c_1\ln(\bar{r}R_0/a)$. 

Since the outer solution diverges logarithmically as $\bar{r}\rightarrow0$, the inner solution must have a switchback logarithmic decomposition. We thus introduce the general form: $W_{\textrm{in}}(R) = f_0(R)+f_1(R)\ln(\varepsilon)+O(\varepsilon)$, where $f_0$ and $f_1$ are two unknown functions. Invoking Eq.~(\ref{innerODE}), the latter satisfy: 
\begin{align}
		\frac{\textrm{d}^2f_0}{\textrm{d} R^2}+\frac{1}{R}\frac{\textrm{d} f_0}{\textrm{d} R} &= -\frac{1}{(1+R^2)^2} ~, \\
		\frac{\textrm{d}^2f_1}{\textrm{d} R^2}+\frac{1}{R}\frac{\textrm{d} f_1}{\textrm{d} R} &= 0~.
\end{align}
Using the boundary conditions at $R = 0$, i.e. $\textrm{d}f_0/\textrm{d}R = 0$ and $\textrm{d}f_1/\textrm{d}R = 0$, one gets:
\begin{align}
	f_0 &= -\frac{1}{4}\ln(1+R^2)+c_3~,\\
	f_1 &= c_4~,
\end{align}
where $c_3$ and $c_4$ are two unknown constants. 

We now proceed to the asymptotic matching of the inner and outer solutions, in order to find the three unknown constants and have an approximate expression of the membrane deflection over the full spatial range. The matching condition reads:
\begin{align}
	f_0|_{R\rightarrow+\infty}+f_1|_{R\rightarrow+\infty}\ln(\varepsilon) &= \bar{w}_{\textrm{out}}|_{r\rightarrow0}~, \\
	\Rightarrow    -\frac{1}{2}\ln\left(\frac{\bar{r}}{\sqrt{2\epsilon}}\right)+c_3+c_4\ln(\varepsilon) &= c_1\ln\left(\frac{R_0\bar{r}}{a}\right)~,
\end{align}
which results in $c_1 = -1/2$, $c_3 = \ln\left[a/\left(R_0\sqrt{2}\right)\right]/2$, and $c_4 = -1/4$. As a consequence, the solution $W(R)=\bar{w}(\bar{r})=w(r)/w^*$ over the full spatial range can be approximated by the matched expression:
\begin{align}
	W &= W_{\textrm{out}}+W_{\textrm{in}}-W_{\textrm{in}}|_{R\rightarrow+\infty},\\
	 &= \frac{1}{4}\ln\left[\frac{a^2}{2\epsilon R_0^2(1+R^2)}\right]\ . \label{eq:pre_stress_response}
\end{align}
Putting back dimensions, one finally gets:
\begin{align}
 w(r) = \frac{\epsilon_0U^2R_0}{4D\sigma_0t_{\textrm{m}}}\ln\left[\frac{a^2}{2DR_0+r^2}\right]~.
\label{eq:pre_stress_response_full}
\end{align}
Thus, the deformation at the center ($r=0$) of the membrane reads:
\begin{align}
 w_0 = \frac{\epsilon_0U^2R_0}{4D\sigma_0t_{\textrm{m}}}\ln\left[\frac{a^2}{2DR_0}\right]~.
\label{eq:pre_stress_response_full}
\end{align}

\begin{figure}[ht!]
    \centering
    \includegraphics[width=0.4\textwidth]{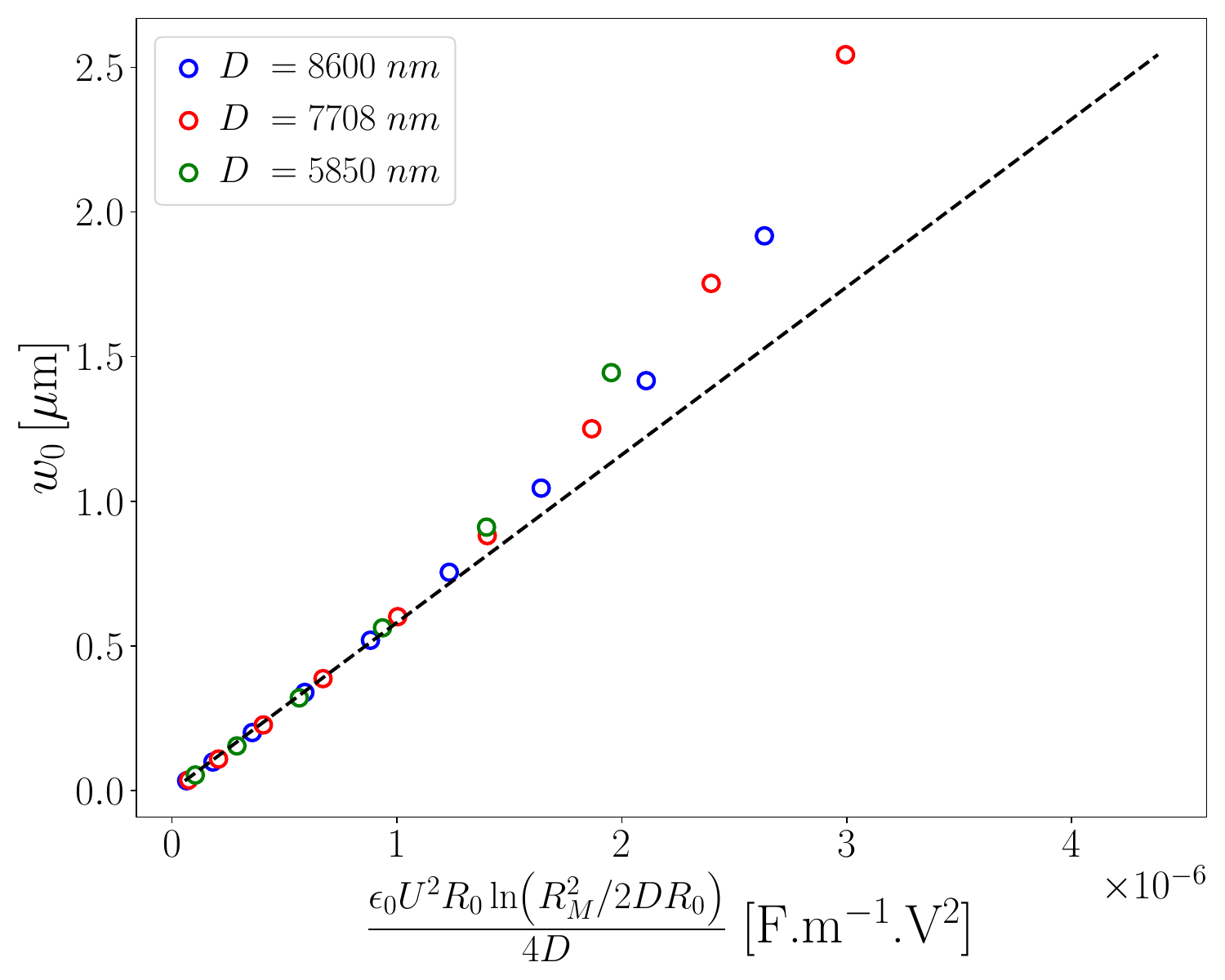}
    \caption{Measured central deflection $w_0$ of the membrane, as a function of the squared voltage $U^2$ re-scaled according to Eq.~(\ref{eq:pre_stress_response_full}), for different gap distances $D$, as indicated. The dashed black line is a fit of the data for $w_0<1~ \mu$m, with $N_0= 1.67$~N/m as the single fit parameter (equal to the inverse of the slope of the linear fit). $t_m = 13.8~ \mu$m, T = 100$^\circ$C, $X=1{:}5$.}
    \label{deltavsVfit}
\end{figure}

The central deflection measured for a single membrane at three different gap distances is shown as a function of the rescaled squared voltage in Fig.~\ref{deltavsVfit}. As can be seen, Eq.~(\ref{eq:pre_stress_response_full}) accurately represents the data at low membrane deflections ($w_0<1~ \mu$m), and can be used to extract the tension of the membrane $N_0=1.67$~N/m. Conversely, if the tension on the membrane was known beforehand, which can calculated from its resonance frequency, an accurate determination of the applied force field could be achieved with this method. At larger membrane deflections the weakly deformed approximation (Eq.~(\ref{eq:geometry})) is no longer valid. Let us end the discussion with a remark on nonlinearities in the deflection, that we neglected. Based on previous calculations,~\cite{campbell1956theory} nonlinearities would lead to a deflection that would be proportional to $p^{1/3}$. Since the latter behavior is not observed in our experimental results, we conclude that nonlinear aspects remain insignificant within our parametric range and that pre-stress dominates.

\section*{Conclusion}
We have described the development, principles, and calibration of a novel class of Surface Forces Apparatus (SFA): the Soft SFA. It involves a compliant elastomeric membrane instead of the rigid surfaces used in classical SFA. The interest of this configuration is threefold. First, the membrane being compliant, it can serve as a force probe itself and no external spring is required for force measurement purposes. Secondly, its large compliance allows for the improvement in  force sensibility, as compared to classical SFAs. The nanonewton range is already reached with the current window of parameters, and could still be improved further. Lastly, the setting is an ideal platform to investigate the intricate coupling between confined nanofluidics and complex soft interfaces, which is ubiquitous in material science, nanophysics and biophysics.

\begin{acknowledgments}
The authors thank Etienne Barthel for interesting discussions. They acknowledge financial support from the Agence Nationale de la Recherche under
  Softer (ANR21-CE06-0029) and Fricolas (ANR-21-CE06-0039) grants, as well as from the Interdisciplinary and 
 Exploratory Research Program under a MISTIC grant at the University of Bordeaux, France. 
The authors also acknowledge financial support from the European Union through the European Research Council under EMetBrown (ERC-CoG-101039103) grant. Views and opinions expressed are however those of the authors only and do not necessarily reflect those of the European Union or the European Research Council. Neither the European Union nor the granting authority can be held 
 responsible for them. Finally, they thank the RRI Frontiers of Life, which 
  received financial support from the French government in the framework of the University of Bordeaux's France 2030 program,
   as well as the Soft Matter Collaborative Research Unit, Frontier Research Center for Advanced Material and Life Science, 
 Faculty of Advanced Life Science, Hokkaido University, Sapporo, Japan; and the CNRS International Research Network between France and India on ``Hydrodynamics at small scales: from soft matter to bioengineering".
\end{acknowledgments}

\section*{Data Availability }
The data supporting the findings of this study are available upon reasonable request to the authors.
\nocite{*}
\bibliography{aipsamp}
\end{document}